 \documentclass[aps,prx,groupedaddress,nofootinbib,reprint]{revtex4-1}
\usepackage{times}
\usepackage{bm}
\usepackage{graphicx}
\usepackage{color}
\usepackage[usenames,dvipsnames]{xcolor}
\usepackage{ulem}

\DeclareMathAlphabet{\mathscr}{OT1}{pzc}{m}{it}

\newcommand{\expect}[1]{{\left\langle #1 \right\rangle}}

\newcommand{\noopsort}[1]{}

\def \expect#1{{\left \langle #1 \right\rangle}}

\newcommand{\be}{\begin{eqnarray}}
\newcommand{\ee}{\end{eqnarray}}

\def\Dan#1{{\textcolor{ForestGreen}{#1}}}

\def\Xe#1{{$^{#1}$Xe}}

\begin{document}

 \title{Synchronous Spin-Exchange Optical Pumping}

\def\Wisc{Department of Physics, University of Wisconsin-Madison, Madison, WI 53706, USA}
\author{A. Korver}
\author{D. Thrasher}
\author{M. Bulatowicz}
\author{T. G. Walker}
\affiliation{\Wisc}

\date{\today}

\begin{abstract} 

{We describe a new approach to precision NMR with hyperpolarized gases designed to mitigate NMR frequency shifts due to the alkali spin exchange field.
The electronic spin polarization of optically pumped alkali atoms is square-wave modulated at the noble-gas NMR frequency and oriented transverse to the DC Fourier component of the NMR bias field.  Noble gas NMR is driven by spin-exchange collisions with the oscillating electron spins. On resonance, the time-average torque from the oscillating spin-exchange field produced by the alkali spins is zero.  Implementing the NMR bias field as a sequence of alkali 2$\pi$-pulses enables synchronization of the alkali and noble gas spins despite a $1000$-fold discrepancy in gyromagnetic ratio. We demonstrate this method with Rb and Xe, and observe novel NMR broadening effects due to the transverse oscillating spin exchange field. When uncompensated, the spin-exchange field at high density broadens the NMR linewidth by an order of magnitude, with an even more dramatic suppression (up to 70x) of the phase shift between the precessing alkali and Xe polarizations.  When we introduce a transverse compensation field, we are able to eliminate the spin-exchange broadening and restore the usual NMR phase sensitivity. 
{The projected quantum-limited sensitivity is better than 1 nHz/$/\sqrt{\rm Hz}$. }
}


\end{abstract}

\maketitle


The ability to produce highly magnetized noble gases via spin-exchange collisions with spin-polarized alkali atoms \cite{WalkerRMP} has greatly impacted scientific studies of magnetic resonance imaging \cite{Walkup2014}, high-energy nuclear physics with spin-polarized targets \cite{Singh2015}, and chemical physics \cite{Jimenez-Martinez2014}.  Applications in precision measurements began with NMR gyros \cite{Meyer2014} and have continued with fundamental symmetry tests using multiple cell free induction decay \cite{Vold1984}, dual-species masers \cite{Glenday2008,Rosenberry2001}, self-compensating co-magnetometers \cite{Smiciklas2011},  NMR oscillators \cite{Yoshimi2012}, and free spin-precession co-magnetometers  \cite{Bulatowicz2013,Tullney2013,Allmendinger2014,Sheng2014}.  

Some of these approaches \cite{Meyer2014,Smiciklas2011,Yoshimi2012,Bulatowicz2013,Sheng2014} take advantage of enhanced NMR detection by the embedded alkali magnetometer.  The alkali and noble-gas spin ensembles experience enhanced  polarization sensitivity due to the Fermi-contact interaction during collisions between the two species.  The effective Fermi-contact fields experienced by the two species are
\be
{\bm B}_{SK}&=&b_{SK}\expect{\bm K}={8\pi\kappa\over 3}{\mu_K\over K}n_K \expect{\bm K}
\label{BSK}\\
{\bm B}_{KS}&=&b_{KS}\expect{\bm S}=-{8\pi\kappa\over 3} g_S\mu_B n_S \expect{\bm S}\label{BKS}
\ee
where ${\bm S}, {\bm K}$ are the electron and nuclear spin operators, $n_S,n_K$ the atomic densities, $\mu_K$ the nuclear magnetic moment of the noble gas, $\mu_B$ the Bohr magneton, and the atomic g-factor $g\approx 2$.
The frequency-shift enhancement factor $\kappa$ \cite{Grover78,Schaefer89} was recently measured to be $493\pm31$ \cite{Ma2011} for RbXe.  Thus the detected NMR field ${\bm B}_{SK}$ is $\sim$500$\times$ larger for the embedded magnetometer than for any external sensor.  This seemingly decisive advantage comes with the cost of similarly enhancing the  ${\bm B}_{KS}$ field due to the spin-polarized alkali atoms,  {190 $\mu$G at $2n_S \expect{\bm S}=10^{13}$ cm$^{-3}$}. In typical  longitudinally polarized NMR this field produces large frequency shifts of order 0.1 Hz.  One approach for mitigating this effect is to compare two Xe isotopes \cite{Bulatowicz2013,Meyer2014}, for which the enhancement factors are equal to about 0.1\% \cite{Bulatowicz2013}.  Another recent strategy nulls the alkali field by saturating the alkali electron spin resonance during free precession intervals  \cite{Sheng2014}.

\begin{figure}[bht]
\includegraphics[width=3.5 in]{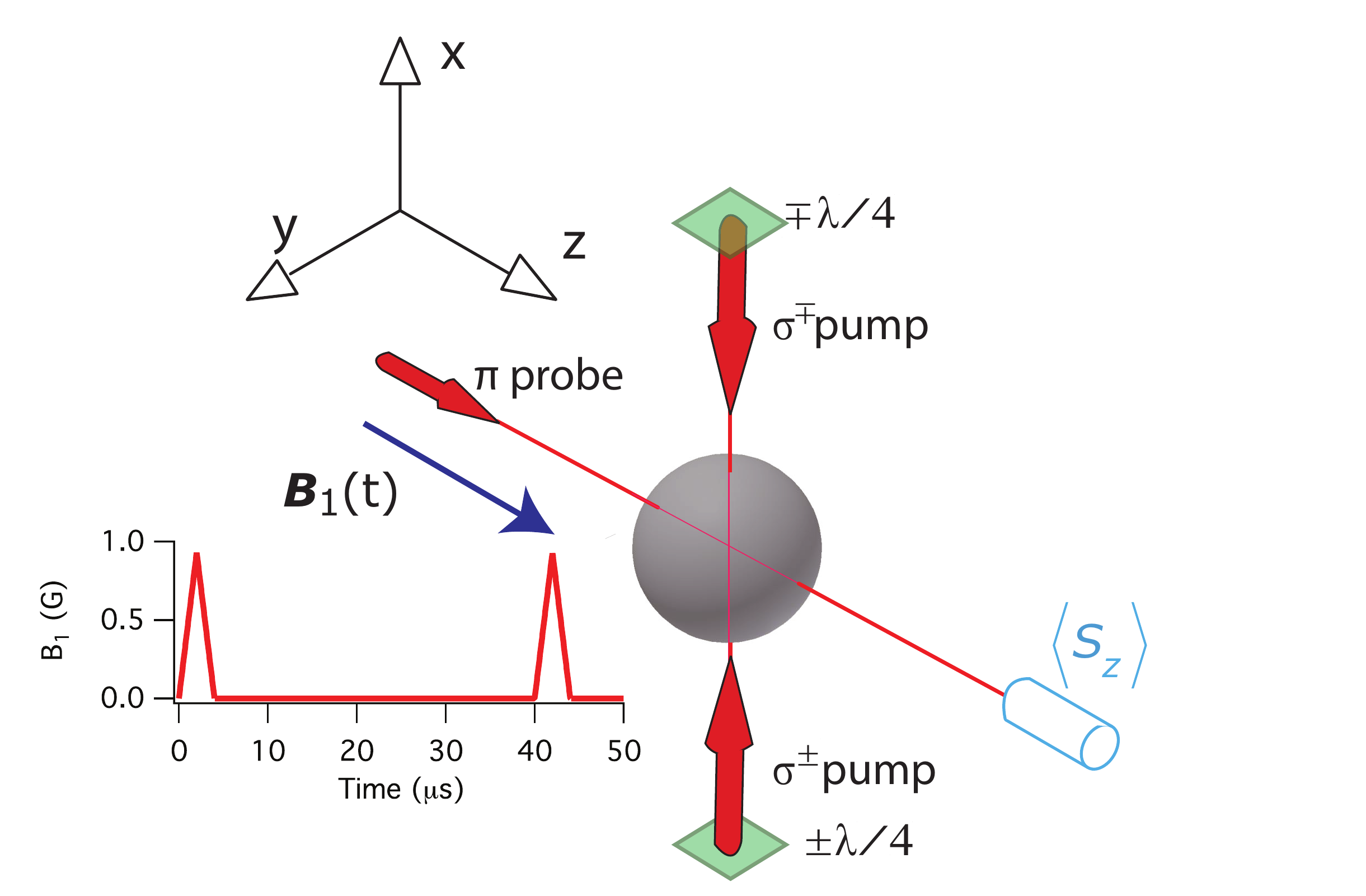}
\caption{Essential components of synchronous spin-exchange optical pumping.  An NMR bias field $\bm B_1(t)$ is applied as a sequence of short alkali 2$\pi$ pulses, allowing the atoms to be optically pumped with resonant light perpendicular to $\bm B_1$ with periodic polarization reversals from the polarization modulators.  Spin-exchange collisions between Xe atoms and the oscillating Rb spins drive 
the NMR resonance.  { The field produced by the precessing Xe nuclei rotates the Rb spins which in turn cause a Faraday rotation of the polarization of the probe light.}}\label{fig:setup}
\end{figure}

{This Letter describes a new approach, synchronous spin-exchange optical pumping, in which both the alkali  and noble gas spins are polarized transverse to the bias magnetic field. 
 The essential components are shown in Fig. ~\ref{fig:setup}. }This bias magnetic field is applied as a sequence of short alkali $2\pi$ pulses, allowing the alkali atoms to be polarized perpendicular to the bias field \cite{Korver2013}. 
By periodically reversing the alkali polarization direction at the noble gas resonance frequency, transverse nuclear polarization is resonantly generated by spin-exchange collisions. {On} resonance, the alkali polarization is in phase with the noble-gas precession, so there is no time-averaged torque from the alkali field and therefore no frequency shift of the NMR resonance. 
However, when the alkali polarization is modulated somewhat off-resonance, we find that the torque from the alkali field suppresses the phase shift of the NMR precession and generates a longitudinal component of the noble-gas polarization.  This results in a novel broadening of the NMR resonance that is readily suppressed by adding an  AC compensation field 180$^\circ$  out of phase with and along the same direction as the Rb polarization modulation.  When fully compensated, the broadening effect becomes negligible and this approach thus maintains the SNR advantages of NMR detection by the embedded alkali magnetometer, while suppressing the Fermi contact field shifts.  This is a promising approach for ultra sensitive NMR using hyper polarized Xe and $^3$He in the future.

{Synchronous spin-exchange is accomplished by a collisional variant of the  Bell-Bloom method of synchronous optical pumping \cite{Bell1961}. }
 The nuclei are polarized by spin-exchange collisions with  alkali-metal  atoms whose spin,  oriented transverse to a bias magnetic field ${\bm B}_1=B_1(t) \hat{z}$,  oscillates at the nuclear Larmor frequency.
  The combined effects of Larmor precession, transverse spin-relaxation, and spin-exchange collisions are described by the Bloch equation for the transverse spin components $\langle K_+\rangle=\expect{K_x}+i \expect{K_y}$:
\be
{d\over dt}\langle K_+\rangle=-(i\gamma_K \bar B_1+\Gamma_2)\langle K_+\rangle+\Gamma_{\rm SE}\langle S_x\rangle(t)
\label{SSEOPeqn}\ee
The noble-gas gyromagnetic ratio is $\gamma_K$, the transverse relaxation rate of the nuclei is $\Gamma_2$, and the spin-exchange collision rate is $\Gamma_{\rm SE}$.  A notable omission from Eq.~\ref{SSEOPeqn} is the torque from the magnetic field $\hat{x}b_{KS}\langle S_x\rangle(t)$
 produced by the alkali spins, which we temporarily assume has been eliminated as will be explained in detail below.

The alkali spin-polarization $\langle S_x\rangle(t)=\sum_q s_q e^{-i q \omega t}$ is modulated at a frequency $\omega$ near the $p$-th submultiple of the time-averaged noble-gas resonance frequency $\omega_0=\gamma_K\bar B_1$.  In the rotating-wave approximation we neglect all but the near-resonant co-rotating Fourier component $s_p$, which produces a transverse noble-gas polarization
\be
\langle K_+\rangle={\Gamma_{\rm SE} s_p e^{-i p\omega t}\over \Gamma_2+i(\omega_0-p\omega)}=\langle \tilde K_+\rangle e^{-i p\omega t}
\label{ideal}
\ee
with the usual amplitude and phase response of a driven oscillator. {On resonance, the transverse polarization in the rotating frame} is $K_{\rm max}= \Gamma_{\rm SE}s_p/\Gamma_2$.  Off-resonance, the rotating frame magnetic field $\hat{z}(\omega_0-p\omega)/\gamma_K$ causes the polarization to tilt in the x-y plane by an angle $\delta\phi={\tan^{-1}}[(\omega_0-p\omega)/ \Gamma_2]$\Dan{.}
In comparison to the usual longitudinal spin-exchange \cite{WalkerRMP}, the maximum polarization attainable depends on a competition between $\Gamma_{\rm SE}$ and $\Gamma_2$ rather than the longitudinal relaxation rate $\Gamma_1$.  The polarization is also reduced by the Fourier amplitude $|s_p|/|\langle S_x\rangle |\approx {2/\pi p}$ for our roughly square wave polarization modulation. When combined with a feedback system to actively null the phase shift $\delta\phi$, we observe that the resonance frequency is an accurate measure of the Larmor frequency: $\omega=\omega_0/p$.  Thus synchronous spin-exchange is attractive for accurate absolute magnetometry and rotation sensing.

The synchronously oscillating transverse alkali polarization, an impossibility in a DC magnetic field due to the $\sim 1000$ fold larger magnetic moment as compared to Xe, is enabled by replacing the usual DC bias field by $B_1(t)$, a sequence of  short (4 $\mu$s) pulses.  Each pulse  produces $2\pi$ precession of the alkali atoms \cite{Korver2013}, {\it i.e.} $\int {\gamma_{Rb}}B_1dt=2\pi$.  Here the Rb gyromagnetic ratio is $\gamma_{Rb}=\gamma_S/(2I+1)\approx  \mu_B/3\hbar$ for the $^{85}$Rb isotope we use.  The application of a sequence of 2$\pi$ pulses causes no time-averaged precession of the Rb spins, thereby allowing the Rb to be optically pumped as if in zero field.  Meanwhile, the polarized Xe nuclei precess only $2\pi\gamma_{Xe}/\gamma_{Rb}\approx 5$ mrad per pulse (for \Xe{131}) and effectively see only the DC average $\bar B_1$.

\begin{figure}[tb]
\includegraphics[width=3.5 in]{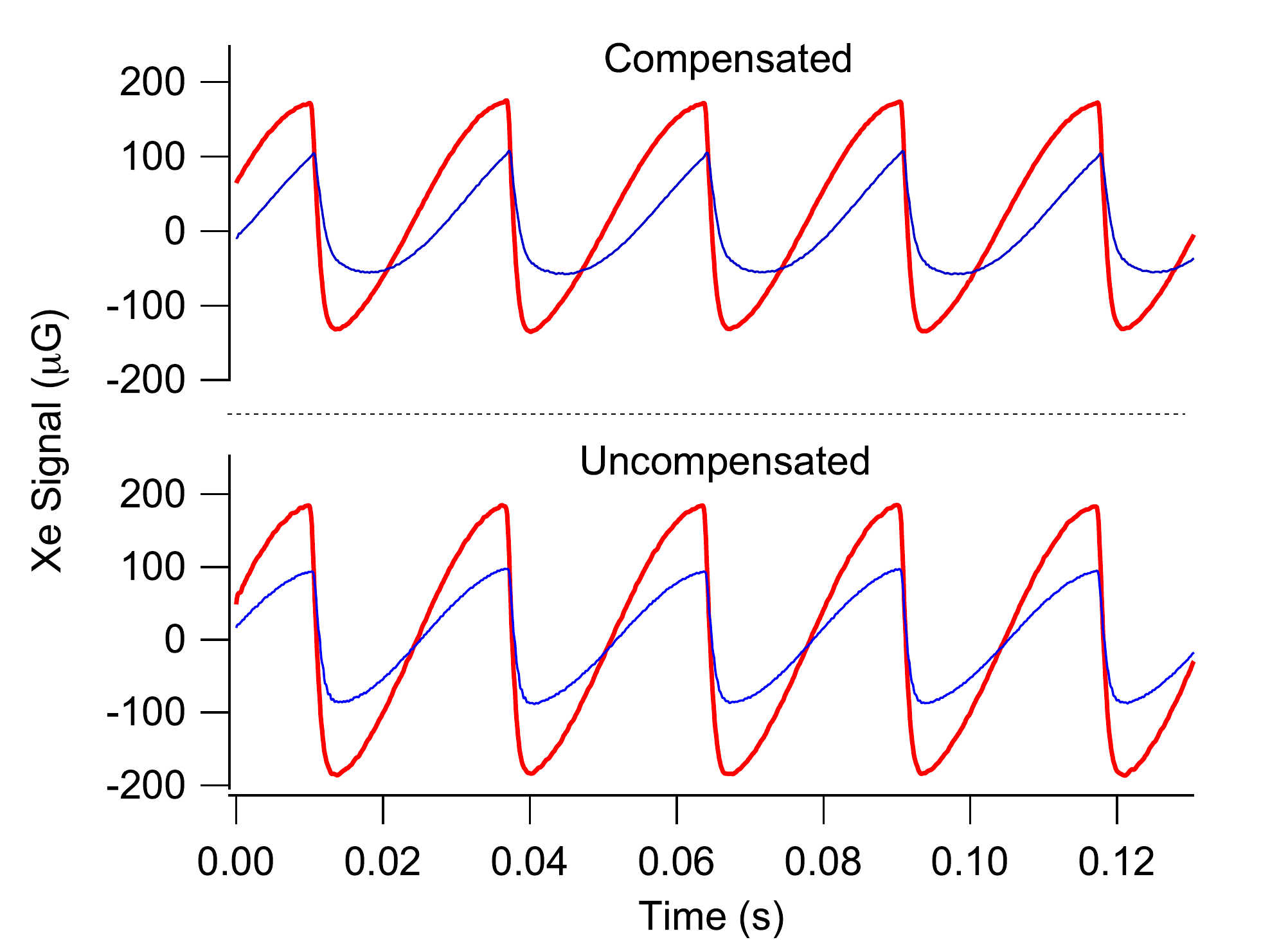}
\caption{ Xe precession signals detected by the embedded Rb magnetometer.  The magnetometer gain reversals square-wave modulate the sinusoidal Xe precession. { When the Rb spin-exchange field is compensated,   signals detuned by one half-linewidth (blue) are  smaller in amplitude and phase-shifted with respect to the on-resonance case (red).  Uncompensated, the spin-exchange fields  greatly suppress phase shifts.}
 }\label{fig:signal}
\end{figure}

{The apparatus includes a nominally spherical 8 mm diameter glass cell that contains Rb vapor} (density $2-20\times 10^{12}$ cm$^{-3}$), 32 Torr of  \Xe{131} gas, 4 Torr of \Xe{129} gas, and 300 Torr of N$_2$. {The cell is held inside a ceramic oven that is heated by running {20 kHz} AC current through commercial Kapton flex-circuit heater strips that are configured to minimize stray fields. }  Outside the oven are Helmholtz field coils for fine-shimming of the 3 magnetic field components, plus a set of 4 coils for generating the $2\pi$ pulses.  The oven and coils are inside a 3 layer magnetic shield with optical access ports for the lasers.  The 2$\pi$ coils are designed to maximize uniformity,  maintain low inductance, and minimize stray fields that couple to the magnetic shields.  They are driven by a custom MOSFET circuit designed to minimize pulse-to-pulse charge fluctuations.

Optical pumping is performed using two free-running {40mW} 795 nm distributed feedback diode lasers that are combined on a non-polarizing beamsplitter. { One of the lasers is tuned above the Rb D1 resonance, the other below.   Since  the AC Stark field changes sign at resonance, we can reduce the effective field from about 2 mG (for a single pumping frequency)  to $<10$ $\mu$G   by adjusting the relative detunings and intensities of the two lasers. }
The two output beams from the polarizing beamsplitter, each containing both pump frequencies, enter the cell and propagate along along the $\pm \hat x$ directions; {doing this substantially improves the uniformity of the optical pumping of the optically thick alkali vapor}. The last optical elements before the cell are wave plates and  liquid crystal variable retarders that allow the circular light polarizations to be reversed in about 200 $\mu$s.  At low Rb density, the laser detunings are selected to produce about 70\% Rb polarization while nulling the AC Stark shifts.  At high densities the polarization drops to about 40\% without significantly degrading the AC Stark cancellation.

An off-resonant 30 mW probe laser, propagating along $\hat{z}$, serves to observe the z-component of the alkali polarization using low-noise Faraday rotation detection.  Because the alkali atoms are optically pumped along $\hat{x}$, the 
Faraday rotation detects the $\hat{y}$-component of the Xe field: $\expect{S_z}=-\gamma_Sb_{SK}\expect{K_y}\expect{S}_x(t)/\Gamma'$.   Note that the magnetometer sensitivity reverses sign as the alkali spin direction is reversed due to the synchronous pumping.  In order to detect $\expect{K_z}$,  we add a $\bm B_2=$1 mG $\hat{x}\sin(\omega_2 t)$  field at $\omega_2=  900$ Hz. Demodulation of $\expect{S_z}$ at $\omega_2$ then gives an output equal to $-\gamma_S B_2 b_{SK}\expect{K_z}\expect{S}_x(t)/\Gamma'^2$.  Characterization of the magnetic response implies a relaxation rate $\Gamma'=80000$/s with a magnetic response bandwidth of $1.5$ kHz and a dynamic range of $5.7$ mG.  The noise floor of the magnetometer is typically 1 nG/$\sqrt{\rm Hz}$, limited by photon-shot-noise.

\begin{figure}[htb]
\includegraphics[width=3.5 in]{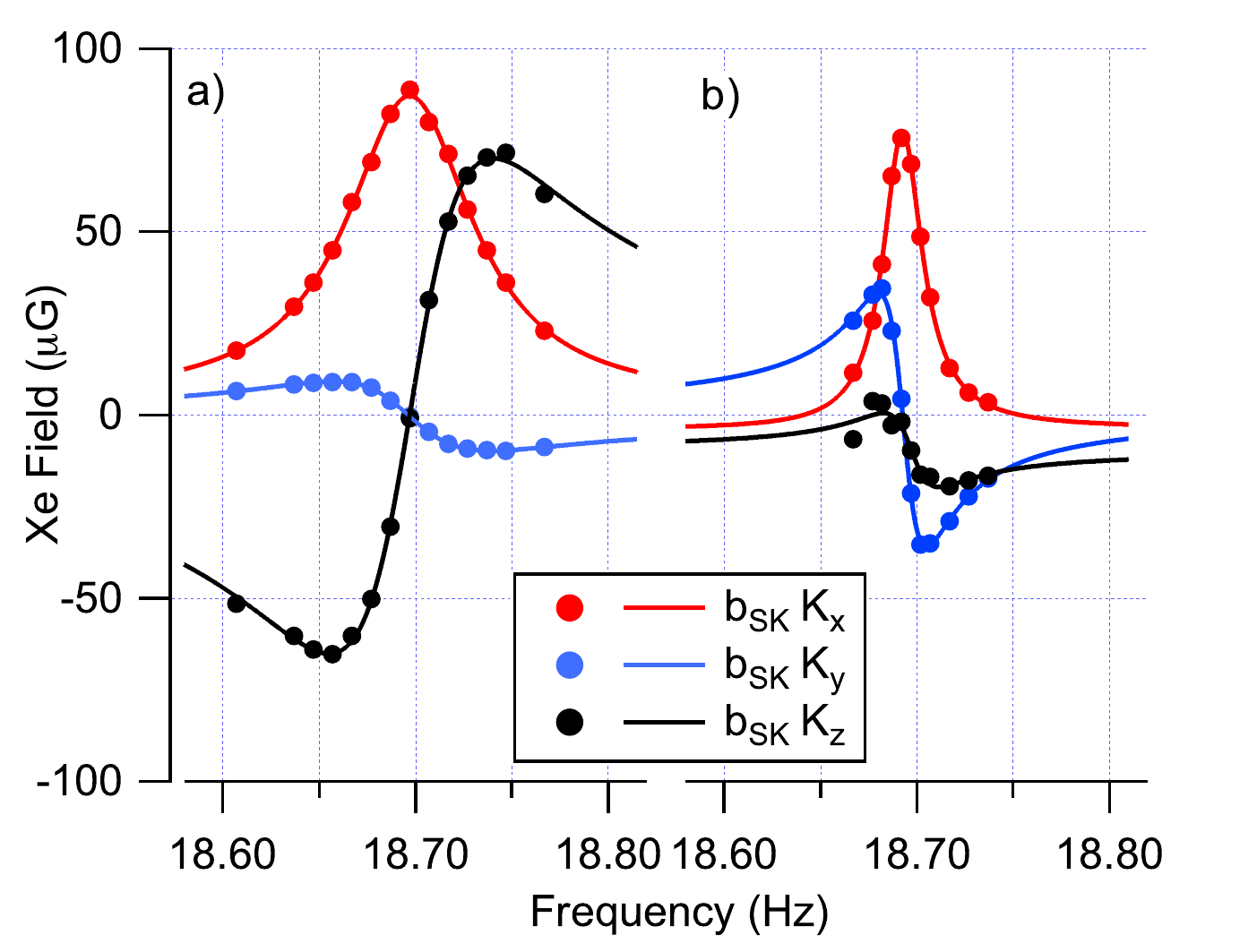}
\caption{NMR line shapes for synchronously pumped \Xe{131}, $n_S=7.1\times 10^{12}$ cm$^{-3}$. Right:  when the spin-exchange field is compensated, the rotating-frame polarizations show the expected Lorentzian forms (red: in phase or $\tilde K_x$, blue: $\tilde K_y$, black: $K_z$), and little longitudinal polarization is generated.  Left:  An uncompensated spin-exchange field not only broadens the NMR resonance, but produces large z-polarizations and dramatically suppresses the rotating frame quadrature component.  }\label{ResonanceCurves}
\end{figure}

To null stray DC magnetic fields, we temporarily turn off the 2$\pi$ pulses while retaining the pump modulation, effectively running the magnetometer as a zero-field magnetometer with polarization modulation.  
After field nulling with the zero-field magnetometer, we turn on the 2$\pi$ pulses. The area of the 2$\pi$ pulses is set to better than $0.1$\% by nulling the magnetometer signal generated by $\bm B_2$.
  The 25 kHz repetition rate of the 2$\pi$ pulse sequence and the modulation waveform for the optical pumping are generated by direct digital synthesis using an FPGA phase-locked to a commercial atomic Rb clock.  
With stray fields nulled and the 2$\pi$ pulse area set, the synchronously pumped Xe signals are readily observed when the pumping modulation frequency $\omega$ is brought near either the $\omega_0=2\pi\times  18.702$ Hz \Xe{131} resonance frequency  or the $\omega_0=2 \pi \times 21.03 $ Hz 3rd subharmonic of the \Xe{129} resonance frequency.  A magnetometer waveform is shown in Fig.~\ref{fig:signal}, and a sample spin-exchange resonance curve is shown in Fig.~\ref{ResonanceCurves}b.  The resonance curve is taken by lock-in detection of the 2$\omega$ magnetometer signal. The in-phase lock-in output is proportional to $\tilde K_x$ while the quadrature is proportional to $\tilde K_y$. The signal amplitude, typically 180 $\mu$G for both isotopes at $n_S=2.5\times 10^{13}$ cm$^{-3}$ when driven at $p=1$,  is consistent with expectations from independent measurements of $\Gamma_1$, $\Gamma_{\rm SE}$, and $S_x$.

So far we have ignored possible effects from the alkali field;  indeed the resonance in Fig.~\ref{ResonanceCurves}b was taken with a compensated alkali field as will be explained below.  However, when uncompensated the alkali field can have a dramatic effect.  Under the conditions of  Fig.~\ref{ResonanceCurves}a, the in-phase response is broadened by a factor of {4} as compared to Fig.~\ref{ResonanceCurves}b.  Even more dramatic is the observed suppression of the quadrature signal by 10x, while a dispersive z-polarization is acquired.  At higher densities we have observed broadening of up to 10$\times$ and quadrature suppression of 75$\times$ for \Xe{129}.


The broadening from the spin-exchange field can be understood as follows.  First, when the pumping is off-resonance, the phase shift $\delta \phi$ between the nuclear precession and the alkali field produces a DC torque $-\gamma_K {\bm B}_{KS}\times \expect{\bm K}={2\over p\pi}\gamma_K |B_{KS}|K_\perp \sin\delta\phi$ and so {the Xe spin }develops a z-polarization.  The z-polarization, which can be a substantial fraction of the transverse polarization (as illustrated in Fig.~\ref{ResonanceCurves}a), then couples with the alkali field to produce a transverse torque that is 90 degrees out of phase with the pumping.  This can be considered as a negative feedback mechanism that tries to null the relative phase between the Xe precession and the pumping.  The net effect is to generate a large z-polarization and to suppress the phase shift between the pumping and the Xe precession.  Under our conditions the alkali field, when uncompensated, is sufficiently strong to make $|\expect{K_z}|>|\expect{K_x}|$ at large detunings.

The Bloch equation for the z-polarization is
\be
{d\over dt}\expect{K_z}=-\gamma_K b_{x}(t) \expect{K_y}-\Gamma_1\expect{K_z}
\ee
which leads to a steady-state
\be
\expect{K_z}={\gamma_K b_{xp}\over \Gamma_1}K_\perp \sin(\delta\phi)=\tan(\alpha) K_\perp \sin(\delta\phi)
\ee
The effective magnetic field $b_x(t)$ seen by the Xe nuclei is the sum of the alkali field and {an anti-parallel  square-wave modulated external compensation field}: $b_x(t)=b_{KS}\expect{S_x}(t)-B_c(t)$.  The $p$-th Fourier component of this field is $b_{xp}$. { The size of the spin-exchange torque is characterized by the parameter $\tan(\alpha)=\gamma_K b_{xp}/ \Gamma_1$.}

\begin{figure}[hbt]
\includegraphics[width=3.2 in]{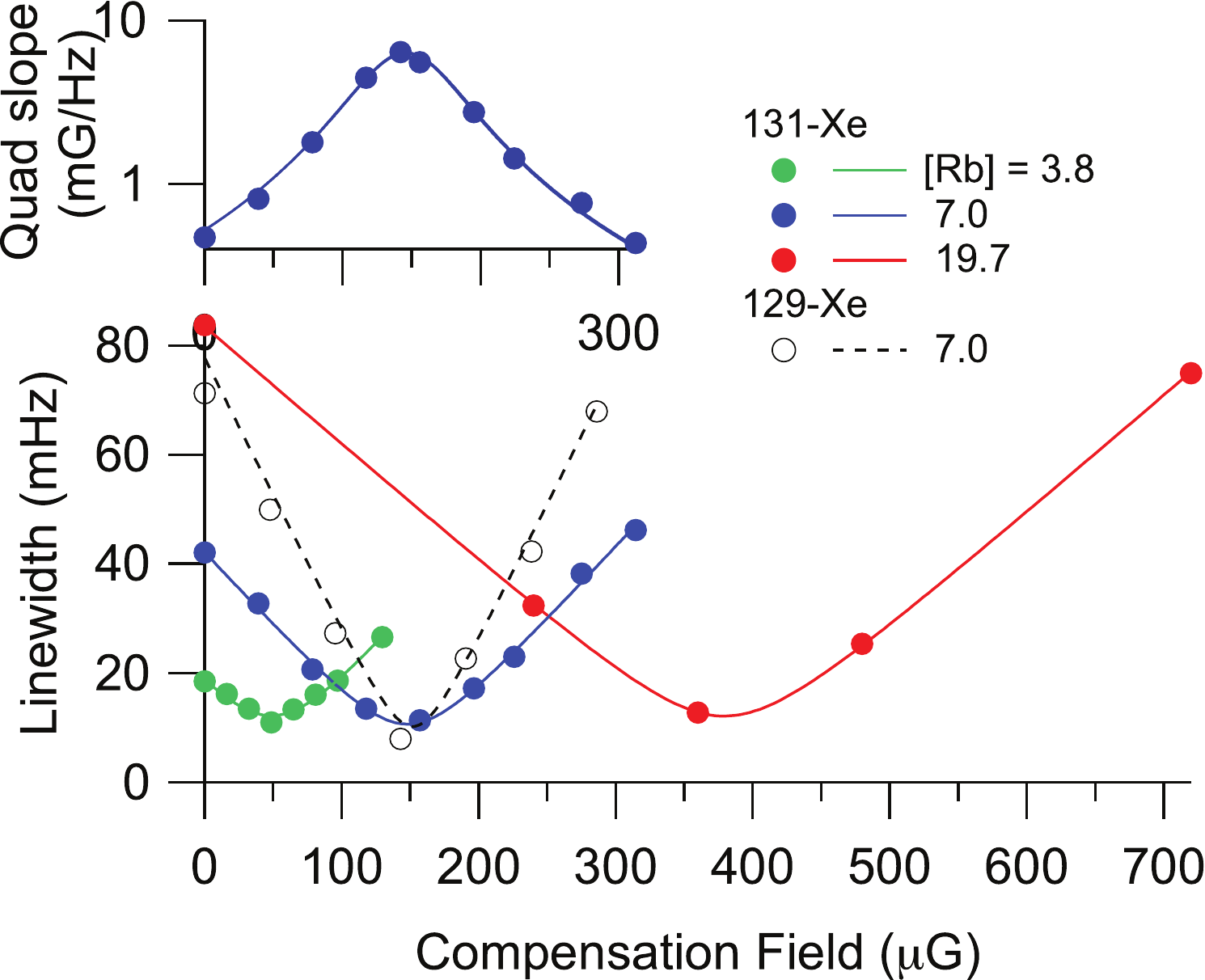}
\caption{Comparison of observed spin-exchange broadening with Eqs.~\ref{Kplus} and \ref{Gammaeff}.  (bottom) The broadening of the NMR linewidth due to the spin-exchange field can be eliminated by applying a compensation field that oscillates 180$^\circ$ out of phase with the spin-exchange field.  Solid lines and dots show \Xe{131} linewidths, dashes and unfilled dots show a sample for \Xe{129}. Densities are listed in units of $10^{12}$ cm$^{-3}$. (top) Representative measurement of the slope of the quadrature signal as a function of compensation field.}\label{Broad}
\end{figure}

The  z-polarization of the Xe now allows the transverse fields to apply a torque, so the transverse components of the Bloch equations become
\be
{d\over dt}\langle K_+\rangle&=&-(i\gamma_K \bar B_1+\Gamma_2)\langle K_+\rangle \nonumber   \\
&&+i\gamma_K b_x{(t)} \expect{K_z}+\Gamma_{\rm SE}\expect{S_x}(t)\\
&\approx&-(i\gamma_K \bar B_1+\Gamma_2)\langle K_+\rangle \nonumber \\
&&\!\!\!+\left(i\Gamma_1 \tan^2(\alpha) \sin(\delta\phi)K_\perp+\Gamma_{\rm SE}s_p\right)e^{-i p \omega t} 
\ee 
where we have again made the rotating wave approximation.  Notice that the torque from the alkali field is 90$^\circ$ out of phase with the pumping torque, acting to suppress the phase shift between the pumping and the precession.  The steady-state polarization in the rotating frame is
\be
\langle \tilde K_+\rangle&=&{K_{\rm max}\over 1+\Delta^2/\Gamma_{\rm eff}^2}\left(1+i {\Gamma_2\Delta\over \Gamma_{\rm eff}^2}\right)  \label{Kplus}\\
\expect{K_z}&=&\tilde \expect{K_y} \tan\alpha
\ee
where the spin-exchange broadened linewidth is
\be
\Gamma_{\rm eff}=\sqrt{\Gamma_2^2+\Gamma_1\Gamma_2 \tan^2\alpha}=\sqrt{\Gamma_2^2+{\Gamma_2\over\Gamma_1}(\gamma_K b_{xp})^2} \label{Gammaeff}
\ee
and the detuning is $\Delta=p \omega-\omega_0$.
For a large spin-exchange field, {\it i.e.} $\tan\alpha\gg1$, these equations account for the broadening, small $\expect{\tilde K_y}$ response, and large $\expect{K_z}$ response.  They also predict that with the application of a compensation field $B_c$ 180$^\circ$ out of phase with the pumping, the linewidth should narrow to $\Gamma_2$, the magnitude of $\expect{\tilde K_y}$ should be restored, and $\expect{K_z}$ should be suppressed, thus bringing the response into agreement with Eq.~\ref{ideal}.  All these features are illustrated in the data of Fig.~\ref{ResonanceCurves}, with fits to Eqs~\ref{Kplus}-\ref{Gammaeff}.

We have studied the spin-exchange broadening over a range of alkali field strengths  for both Xe isotopes.  Figure~\ref{Broad} shows the predicted narrowing of the resonance width with application of the  compensation field.  For Xe-129 there is an order of magnitude narrowing of the width over the field range, despite having reduced $\tan(\alpha)$ by a factor of 3 by pumping at the third subharmonic of the Larmor frequency.  We have confirmed that at a given density the compensation fields for \Xe{129} and \Xe{131} are the same to within {1\%}.
  Figure~\ref{Broad} also shows an example of how the quadrature slope $d\tilde K_y/d\Delta$ dramatically increases as the compensation field approaches the optimum.  We have observed as much as a factor of 75 ratio between the compensated and uncompensated slopes.

At higher densities than reported in this paper, the broadening slows as a function of compensation field.  We believe that this arises because of feedback from the alkali field:  at sufficiently high densities, the tipping of the alkali field off of the $\hat x$-axis becomes important and needs to be taken into account.  Such non-linearities are likely closely related to the studies of strongly coupled alkali and noble gas spin from Ref. \cite{Kornack2002}.

In the future, two-species operation will allow full realization of the spectroscopy potential of synchronous spin-exchange for fundamental symmetry tests.  With two-species, one species can be used to remove magnetic field noise, leaving the other sensitive to non-magnetic interactions.  This strategy is used in various forms in many experiments \cite{Meyer2014,Glenday2008,Rosenberry2001,Bulatowicz2013,Tullney2013,Allmendinger2014,Sheng2014}.  The resulting effective frequency noise $\delta x$ resulting from some non-magnetic interaction $H=h x$ is of order
\be
\delta x=\left(1+{\gamma_2^2\over \gamma_1^2}\right)^{1/2}{\Gamma_2 \delta B\over 2\pi  b_{SK}K_\perp }
\ee 
Our current apparatus is limited by probe photon shot noise at about the $\delta B=10^{-9}$ G/$\sqrt{\rm Hz}$ level, but even this relatively modest magnetometer performance projects to $\delta x=66$ nHz/$\sqrt{\rm Hz}$.  Supposing we can reach the $5\times 10^{-11}$ G/$\sqrt{\rm Hz}$ level of a good magnetic shield, the noise level would be 3 nHz/$\sqrt{\rm Hz}$, competitive with the Ne-Rb-K co-magnetometer of Ref. \cite{Smiciklas2011}.  At the quantum projection noise limit of the magnetometer, the noise level becomes sub-nHz/$\sqrt{\rm Hz}$.  These numbers are very promising for experiments such as sensitive tests of Lorentz violation \cite{Smiciklas2011} and limits on short-range nuclear interactions \cite{Bulatowicz2013,Tullney2013}.

{Finally, and in contrast to the co-magnetometer approach of Ref. \cite{Smiciklas2011}, synchronous spin-exchange promises not only sensitivity but accuracy.  Configured as a dual-species oscillator \cite{Meyer2014} to actively null $\delta\phi$, for example, the nuclear phase becomes an accurate integral of the inertial rotation rate.  The above noise estimates suggest an achievable sensitivity of better than 100 $\mu$deg/$\sqrt{\rm hr}$ for a synchronously pumped NMR gyro with a unity scale factor and a bandwidth that approaches the oscillation frequency.}

We are grateful for many discussions with M. Larsen, R. Wyllie, B. Lancor, M. Ebert, and I. Sulai.   This work was supported by NSF GOALI PHY1306880 and Northrop Grumman Corp.


\begin{thebibliography}{20}%
\makeatletter
\providecommand \@ifxundefined [1]{%
 \@ifx{#1\undefined}
}%
\providecommand \@ifnum [1]{%
 \ifnum #1\expandafter \@firstoftwo
 \else \expandafter \@secondoftwo
 \fi
}%
\providecommand \@ifx [1]{%
 \ifx #1\expandafter \@firstoftwo
 \else \expandafter \@secondoftwo
 \fi
}%
\providecommand \natexlab [1]{#1}%
\providecommand \enquote  [1]{``#1''}%
\providecommand \bibnamefont  [1]{#1}%
\providecommand \bibfnamefont [1]{#1}%
\providecommand \citenamefont [1]{#1}%
\providecommand \href@noop [0]{\@secondoftwo}%
\providecommand \href [0]{\begingroup \@sanitize@url \@href}%
\providecommand \@href[1]{\@@startlink{#1}\@@href}%
\providecommand \@@href[1]{\endgroup#1\@@endlink}%
\providecommand \@sanitize@url [0]{\catcode `\\12\catcode `\$12\catcode
  `\&12\catcode `\#12\catcode `\^12\catcode `\_12\catcode `\%12\relax}%
\providecommand \@@startlink[1]{}%
\providecommand \@@endlink[0]{}%
\providecommand \url  [0]{\begingroup\@sanitize@url \@url }%
\providecommand \@url [1]{\endgroup\@href {#1}{\urlprefix }}%
\providecommand \urlprefix  [0]{URL }%
\providecommand \Eprint [0]{\href }%
\providecommand \doibase [0]{http://dx.doi.org/}%
\providecommand \selectlanguage [0]{\@gobble}%
\providecommand \bibinfo  [0]{\@secondoftwo}%
\providecommand \bibfield  [0]{\@secondoftwo}%
\providecommand \translation [1]{[#1]}%
\providecommand \BibitemOpen [0]{}%
\providecommand \bibitemStop [0]{}%
\providecommand \bibitemNoStop [0]{.\EOS\space}%
\providecommand \EOS [0]{\spacefactor3000\relax}%
\providecommand \BibitemShut  [1]{\csname bibitem#1\endcsname}%
\let\auto@bib@innerbib\@empty
\bibitem [{\citenamefont {Walker}\ and\ \citenamefont
  {Happer}(1997)}]{WalkerRMP}%
  \BibitemOpen
  \bibfield  {author} {\bibinfo {author} {\bibfnamefont {T.~G.}\ \bibnamefont
  {Walker}}\ and\ \bibinfo {author} {\bibfnamefont {W.}~\bibnamefont
  {Happer}},\ }\href@noop {} {\bibfield  {journal} {\bibinfo  {journal} {Rev.
  Mod. Phys.}\ }\textbf {\bibinfo {volume} {69}},\ \bibinfo {pages} {629}
  (\bibinfo {year} {1997})}\BibitemShut {NoStop}%
\bibitem [{\citenamefont {Walkup}\ and\ \citenamefont
  {Woods}(2014)}]{Walkup2014}%
  \BibitemOpen
  \bibfield  {author} {\bibinfo {author} {\bibfnamefont {L.~L.}\ \bibnamefont
  {Walkup}}\ and\ \bibinfo {author} {\bibfnamefont {J.~C.}\ \bibnamefont
  {Woods}},\ }\href@noop {} {\bibfield  {journal} {\bibinfo  {journal} {NMR
  Biomed.}\ }\textbf {\bibinfo {volume} {27}},\ \bibinfo {pages} {1429}
  (\bibinfo {year} {2014})}\BibitemShut {NoStop}%
\bibitem [{\citenamefont {Singh}\ \emph {et~al.}(2015)\citenamefont {Singh},
  \citenamefont {Dolph}, \citenamefont {Tobias}, \citenamefont {Averett},
  \citenamefont {Kelleher}, \citenamefont {Mooney}, \citenamefont {Nelyubin},
  \citenamefont {Wang}, \citenamefont {Zheng},\ and\ \citenamefont
  {Cates}}]{Singh2015}%
  \BibitemOpen
  \bibfield  {author} {\bibinfo {author} {\bibfnamefont {J.~T.}\ \bibnamefont
  {Singh}}, \bibinfo {author} {\bibfnamefont {P.~A.~M.}\ \bibnamefont {Dolph}},
  \bibinfo {author} {\bibfnamefont {W.~A.}\ \bibnamefont {Tobias}}, \bibinfo
  {author} {\bibfnamefont {T.~D.}\ \bibnamefont {Averett}}, \bibinfo {author}
  {\bibfnamefont {A.}~\bibnamefont {Kelleher}}, \bibinfo {author}
  {\bibfnamefont {K.~E.}\ \bibnamefont {Mooney}}, \bibinfo {author}
  {\bibfnamefont {V.~V.}\ \bibnamefont {Nelyubin}}, \bibinfo {author}
  {\bibfnamefont {Y.}~\bibnamefont {Wang}}, \bibinfo {author} {\bibfnamefont
  {Y.}~\bibnamefont {Zheng}}, \ and\ \bibinfo {author} {\bibfnamefont {G.~D.}\
  \bibnamefont {Cates}},\ }\href {\doibase 10.1103/PhysRevC.91.055205}
  {\bibfield  {journal} {\bibinfo  {journal} {Phys. Rev. C}\ }\textbf {\bibinfo
  {volume} {91}},\ \bibinfo {pages} {055205} (\bibinfo {year}
  {2015})}\BibitemShut {NoStop}%
\bibitem [{\citenamefont {Jim{\'e}nez-Mart{\'\i}nez}\ \emph
  {et~al.}(2014)\citenamefont {Jim{\'e}nez-Mart{\'\i}nez}, \citenamefont
  {Kennedy}, \citenamefont {Rosenbluh}, \citenamefont {Donley}, \citenamefont
  {Knappe}, \citenamefont {Seltzer}, \citenamefont {Ring}, \citenamefont
  {Bajaj},\ and\ \citenamefont {Kitching}}]{Jimenez-Martinez2014}%
  \BibitemOpen
  \bibfield  {author} {\bibinfo {author} {\bibfnamefont {R.}~\bibnamefont
  {Jim{\'e}nez-Mart{\'\i}nez}}, \bibinfo {author} {\bibfnamefont {D.~J.}\
  \bibnamefont {Kennedy}}, \bibinfo {author} {\bibfnamefont {M.}~\bibnamefont
  {Rosenbluh}}, \bibinfo {author} {\bibfnamefont {E.~A.}\ \bibnamefont
  {Donley}}, \bibinfo {author} {\bibfnamefont {S.}~\bibnamefont {Knappe}},
  \bibinfo {author} {\bibfnamefont {S.~J.}\ \bibnamefont {Seltzer}}, \bibinfo
  {author} {\bibfnamefont {H.~L.}\ \bibnamefont {Ring}}, \bibinfo {author}
  {\bibfnamefont {V.~S.}\ \bibnamefont {Bajaj}}, \ and\ \bibinfo {author}
  {\bibfnamefont {J.}~\bibnamefont {Kitching}},\ }\href
  {http://dx.doi.org/10.1038/ncomms4908} {\bibfield  {journal} {\bibinfo
  {journal} {Nat. Commun.}\ }\textbf {\bibinfo {volume} {5}} (\bibinfo {year}
  {2014})}\BibitemShut {NoStop}%
\bibitem [{\citenamefont {Meyer}\ and\ \citenamefont
  {Larsen}(2014)}]{Meyer2014}%
  \BibitemOpen
  \bibfield  {author} {\bibinfo {author} {\bibfnamefont {D.}~\bibnamefont
  {Meyer}}\ and\ \bibinfo {author} {\bibfnamefont {M.}~\bibnamefont {Larsen}},\
  }\href {\doibase 10.1134/S2075108714020060} {\bibfield  {journal} {\bibinfo
  {journal} {Gyroscopy and Navigation}\ }\textbf {\bibinfo {volume} {5}},\
  \bibinfo {pages} {75} (\bibinfo {year} {2014})}\BibitemShut {NoStop}%
\bibitem [{\citenamefont {Vold}\ \emph {et~al.}(1984)\citenamefont {Vold},
  \citenamefont {Raab}, \citenamefont {Heckel},\ and\ \citenamefont
  {Fortson}}]{Vold1984}%
  \BibitemOpen
  \bibfield  {author} {\bibinfo {author} {\bibfnamefont {T.~G.}\ \bibnamefont
  {Vold}}, \bibinfo {author} {\bibfnamefont {F.~J.}\ \bibnamefont {Raab}},
  \bibinfo {author} {\bibfnamefont {B.}~\bibnamefont {Heckel}}, \ and\ \bibinfo
  {author} {\bibfnamefont {E.~N.}\ \bibnamefont {Fortson}},\ }\href {\doibase
  10.1103/PhysRevLett.52.2229} {\bibfield  {journal} {\bibinfo  {journal}
  {Phys. Rev. Lett.}\ }\textbf {\bibinfo {volume} {52}},\ \bibinfo {pages}
  {2229} (\bibinfo {year} {1984})}\BibitemShut {NoStop}%
\bibitem [{\citenamefont {Glenday}\ \emph {et~al.}(2008)\citenamefont
  {Glenday}, \citenamefont {Cramer}, \citenamefont {Phillips},\ and\
  \citenamefont {Walsworth}}]{Glenday2008}%
  \BibitemOpen
  \bibfield  {author} {\bibinfo {author} {\bibfnamefont {A.~G.}\ \bibnamefont
  {Glenday}}, \bibinfo {author} {\bibfnamefont {C.~E.}\ \bibnamefont {Cramer}},
  \bibinfo {author} {\bibfnamefont {D.~F.}\ \bibnamefont {Phillips}}, \ and\
  \bibinfo {author} {\bibfnamefont {R.~L.}\ \bibnamefont {Walsworth}},\ }\href
  {\doibase 10.1103/PhysRevLett.101.261801} {\bibfield  {journal} {\bibinfo
  {journal} {Phys. Rev. Lett.}\ }\textbf {\bibinfo {volume} {101}},\ \bibinfo
  {pages} {261801} (\bibinfo {year} {2008})}\BibitemShut {NoStop}%
\bibitem [{\citenamefont {Rosenberry}\ and\ \citenamefont
  {Chupp}(2001)}]{Rosenberry2001}%
  \BibitemOpen
  \bibfield  {author} {\bibinfo {author} {\bibfnamefont {M.~A.}\ \bibnamefont
  {Rosenberry}}\ and\ \bibinfo {author} {\bibfnamefont {T.~E.}\ \bibnamefont
  {Chupp}},\ }\href {\doibase 10.1103/PhysRevLett.86.22} {\bibfield  {journal}
  {\bibinfo  {journal} {Phys. Rev. Lett.}\ }\textbf {\bibinfo {volume} {86}},\
  \bibinfo {pages} {22} (\bibinfo {year} {2001})}\BibitemShut {NoStop}%
\bibitem [{\citenamefont {Smiciklas}\ \emph {et~al.}(2011)\citenamefont
  {Smiciklas}, \citenamefont {Brown}, \citenamefont {Cheuk}, \citenamefont
  {Smullin},\ and\ \citenamefont {Romalis}}]{Smiciklas2011}%
  \BibitemOpen
  \bibfield  {author} {\bibinfo {author} {\bibfnamefont {M.}~\bibnamefont
  {Smiciklas}}, \bibinfo {author} {\bibfnamefont {J.~M.}\ \bibnamefont
  {Brown}}, \bibinfo {author} {\bibfnamefont {L.~W.}\ \bibnamefont {Cheuk}},
  \bibinfo {author} {\bibfnamefont {S.~J.}\ \bibnamefont {Smullin}}, \ and\
  \bibinfo {author} {\bibfnamefont {M.~V.}\ \bibnamefont {Romalis}},\ }\href
  {\doibase 10.1103/PhysRevLett.107.171604} {\bibfield  {journal} {\bibinfo
  {journal} {Phys. Rev. Lett.}\ }\textbf {\bibinfo {volume} {107}},\ \bibinfo
  {pages} {171604} (\bibinfo {year} {2011})}\BibitemShut {NoStop}%
\bibitem [{\citenamefont {Yoshimi}\ \emph {et~al.}(2012)\citenamefont
  {Yoshimi}, \citenamefont {Inoue}, \citenamefont {Furukawa}, \citenamefont
  {Nanao}, \citenamefont {Suzuki}, \citenamefont {Chikamori}, \citenamefont
  {Tsuchiya}, \citenamefont {Hayashi}, \citenamefont {Uchida}, \citenamefont
  {Hatakeyama}, \citenamefont {Kagami}, \citenamefont {Ichikawa}, \citenamefont
  {Miyatake},\ and\ \citenamefont {Asahi}}]{Yoshimi2012}%
  \BibitemOpen
  \bibfield  {author} {\bibinfo {author} {\bibfnamefont {A.}~\bibnamefont
  {Yoshimi}}, \bibinfo {author} {\bibfnamefont {T.}~\bibnamefont {Inoue}},
  \bibinfo {author} {\bibfnamefont {T.}~\bibnamefont {Furukawa}}, \bibinfo
  {author} {\bibfnamefont {T.}~\bibnamefont {Nanao}}, \bibinfo {author}
  {\bibfnamefont {K.}~\bibnamefont {Suzuki}}, \bibinfo {author} {\bibfnamefont
  {M.}~\bibnamefont {Chikamori}}, \bibinfo {author} {\bibfnamefont
  {M.}~\bibnamefont {Tsuchiya}}, \bibinfo {author} {\bibfnamefont
  {H.}~\bibnamefont {Hayashi}}, \bibinfo {author} {\bibfnamefont
  {M.}~\bibnamefont {Uchida}}, \bibinfo {author} {\bibfnamefont
  {N.}~\bibnamefont {Hatakeyama}}, \bibinfo {author} {\bibfnamefont
  {S.}~\bibnamefont {Kagami}}, \bibinfo {author} {\bibfnamefont
  {Y.}~\bibnamefont {Ichikawa}}, \bibinfo {author} {\bibfnamefont
  {H.}~\bibnamefont {Miyatake}}, \ and\ \bibinfo {author} {\bibfnamefont
  {K.}~\bibnamefont {Asahi}},\ }\href {\doibase
  http://dx.doi.org/10.1016/j.physleta.2012.04.043} {\bibfield  {journal}
  {\bibinfo  {journal} {Phys. Lett. A}\ }\textbf {\bibinfo {volume} {376}},\
  \bibinfo {pages} {1924 } (\bibinfo {year} {2012})}\BibitemShut {NoStop}%
\bibitem [{\citenamefont {Bulatowicz}\ \emph {et~al.}(2013)\citenamefont
  {Bulatowicz}, \citenamefont {Griffith}, \citenamefont {Larsen}, \citenamefont
  {Mirijanian}, \citenamefont {Fu}, \citenamefont {Smith}, \citenamefont
  {Snow}, \citenamefont {Yan},\ and\ \citenamefont {Walker}}]{Bulatowicz2013}%
  \BibitemOpen
  \bibfield  {author} {\bibinfo {author} {\bibfnamefont {M.}~\bibnamefont
  {Bulatowicz}}, \bibinfo {author} {\bibfnamefont {R.}~\bibnamefont
  {Griffith}}, \bibinfo {author} {\bibfnamefont {M.}~\bibnamefont {Larsen}},
  \bibinfo {author} {\bibfnamefont {J.}~\bibnamefont {Mirijanian}}, \bibinfo
  {author} {\bibfnamefont {C.~B.}\ \bibnamefont {Fu}}, \bibinfo {author}
  {\bibfnamefont {E.}~\bibnamefont {Smith}}, \bibinfo {author} {\bibfnamefont
  {W.~M.}\ \bibnamefont {Snow}}, \bibinfo {author} {\bibfnamefont
  {H.}~\bibnamefont {Yan}}, \ and\ \bibinfo {author} {\bibfnamefont {T.~G.}\
  \bibnamefont {Walker}},\ }\href@noop {} {\bibfield  {journal} {\bibinfo
  {journal} {Phys. Rev, Lett.}\ }\textbf {\bibinfo {volume} {{111}}},\ \bibinfo
  {pages} {102001} (\bibinfo {year} {{2013}})}\BibitemShut {NoStop}%
\bibitem [{\citenamefont {Tullney}\ \emph {et~al.}(2013)\citenamefont
  {Tullney}, \citenamefont {Allmendinger}, \citenamefont {Burghoff},
  \citenamefont {Heil}, \citenamefont {Karpuk}, \citenamefont {Kilian},
  \citenamefont {Knappe-Grueneberg}, \citenamefont {Mueller}, \citenamefont
  {Schmidt}, \citenamefont {Schnabel}, \citenamefont {Seifert}, \citenamefont
  {Sobolev},\ and\ \citenamefont {Trahms}}]{Tullney2013}%
  \BibitemOpen
  \bibfield  {author} {\bibinfo {author} {\bibfnamefont {K.}~\bibnamefont
  {Tullney}}, \bibinfo {author} {\bibfnamefont {F.}~\bibnamefont
  {Allmendinger}}, \bibinfo {author} {\bibfnamefont {M.}~\bibnamefont
  {Burghoff}}, \bibinfo {author} {\bibfnamefont {W.}~\bibnamefont {Heil}},
  \bibinfo {author} {\bibfnamefont {S.}~\bibnamefont {Karpuk}}, \bibinfo
  {author} {\bibfnamefont {W.}~\bibnamefont {Kilian}}, \bibinfo {author}
  {\bibfnamefont {S.}~\bibnamefont {Knappe-Grueneberg}}, \bibinfo {author}
  {\bibfnamefont {W.}~\bibnamefont {Mueller}}, \bibinfo {author} {\bibfnamefont
  {U.}~\bibnamefont {Schmidt}}, \bibinfo {author} {\bibfnamefont
  {A.}~\bibnamefont {Schnabel}}, \bibinfo {author} {\bibfnamefont
  {F.}~\bibnamefont {Seifert}}, \bibinfo {author} {\bibfnamefont
  {Y.}~\bibnamefont {Sobolev}}, \ and\ \bibinfo {author} {\bibfnamefont
  {L.}~\bibnamefont {Trahms}},\ }\href@noop {} {\bibfield  {journal} {\bibinfo
  {journal} {Phys Rev. Lett.}\ }\textbf {\bibinfo {volume} {{111}}},\ \bibinfo
  {pages} {100801} (\bibinfo {year} {{2013}})}\BibitemShut {NoStop}%
\bibitem [{\citenamefont {Allmendinger}\ \emph {et~al.}(2014)\citenamefont
  {Allmendinger}, \citenamefont {Heil}, \citenamefont {Karpuk}, \citenamefont
  {Kilian}, \citenamefont {Scharth}, \citenamefont {Schmidt}, \citenamefont
  {Schnabel}, \citenamefont {Sobolev},\ and\ \citenamefont
  {Tullney}}]{Allmendinger2014}%
  \BibitemOpen
  \bibfield  {author} {\bibinfo {author} {\bibfnamefont {F.}~\bibnamefont
  {Allmendinger}}, \bibinfo {author} {\bibfnamefont {W.}~\bibnamefont {Heil}},
  \bibinfo {author} {\bibfnamefont {S.}~\bibnamefont {Karpuk}}, \bibinfo
  {author} {\bibfnamefont {W.}~\bibnamefont {Kilian}}, \bibinfo {author}
  {\bibfnamefont {A.}~\bibnamefont {Scharth}}, \bibinfo {author} {\bibfnamefont
  {U.}~\bibnamefont {Schmidt}}, \bibinfo {author} {\bibfnamefont
  {A.}~\bibnamefont {Schnabel}}, \bibinfo {author} {\bibfnamefont
  {Y.}~\bibnamefont {Sobolev}}, \ and\ \bibinfo {author} {\bibfnamefont
  {K.}~\bibnamefont {Tullney}},\ }\href@noop {} {\bibfield  {journal} {\bibinfo
   {journal} {Phys Rev. Lett.}\ }\textbf {\bibinfo {volume} {{112}}},\ \bibinfo
  {pages} {110801} (\bibinfo {year} {{2014}})}\BibitemShut {NoStop}%
\bibitem [{\citenamefont {Sheng}\ \emph {et~al.}(2014)\citenamefont {Sheng},
  \citenamefont {Kabcenell},\ and\ \citenamefont {Romalis}}]{Sheng2014}%
  \BibitemOpen
  \bibfield  {author} {\bibinfo {author} {\bibfnamefont {D.}~\bibnamefont
  {Sheng}}, \bibinfo {author} {\bibfnamefont {A.}~\bibnamefont {Kabcenell}}, \
  and\ \bibinfo {author} {\bibfnamefont {M.~V.}\ \bibnamefont {Romalis}},\
  }\href@noop {} {\bibfield  {journal} {\bibinfo  {journal} {Phys Rev. Lett.}\
  }\textbf {\bibinfo {volume} {{113}}},\ \bibinfo {pages} {163002} (\bibinfo
  {year} {{2014}})}\BibitemShut {NoStop}%
\bibitem [{\citenamefont {Grover}(1978)}]{Grover78}%
  \BibitemOpen
  \bibfield  {author} {\bibinfo {author} {\bibfnamefont {B.~C.}\ \bibnamefont
  {Grover}},\ }\href@noop {} {\bibfield  {journal} {\bibinfo  {journal} {Phys.
  Rev. Lett.}\ }\textbf {\bibinfo {volume} {40}},\ \bibinfo {pages} {391}
  (\bibinfo {year} {1978})}\BibitemShut {NoStop}%
\bibitem [{\citenamefont {Schaefer}\ \emph {et~al.}(1989)\citenamefont
  {Schaefer}, \citenamefont {Cates}, \citenamefont {Chien}, \citenamefont
  {Gonatas}, \citenamefont {Happer},\ and\ \citenamefont
  {Walker}}]{Schaefer89}%
  \BibitemOpen
  \bibfield  {author} {\bibinfo {author} {\bibfnamefont {S.~R.}\ \bibnamefont
  {Schaefer}}, \bibinfo {author} {\bibfnamefont {G.~D.}\ \bibnamefont {Cates}},
  \bibinfo {author} {\bibfnamefont {T.-R.}\ \bibnamefont {Chien}}, \bibinfo
  {author} {\bibfnamefont {D.}~\bibnamefont {Gonatas}}, \bibinfo {author}
  {\bibfnamefont {W.}~\bibnamefont {Happer}}, \ and\ \bibinfo {author}
  {\bibfnamefont {T.~G.}\ \bibnamefont {Walker}},\ }\href {\doibase
  10.1103/PhysRevA.39.5613} {\bibfield  {journal} {\bibinfo  {journal} {Phys.
  Rev. A}\ }\textbf {\bibinfo {volume} {39}},\ \bibinfo {pages} {5613}
  (\bibinfo {year} {1989})}\BibitemShut {NoStop}%
\bibitem [{\citenamefont {Ma}\ \emph {et~al.}(2011)\citenamefont {Ma},
  \citenamefont {Sorte},\ and\ \citenamefont {Saam}}]{Ma2011}%
  \BibitemOpen
  \bibfield  {author} {\bibinfo {author} {\bibfnamefont {Z.~L.}\ \bibnamefont
  {Ma}}, \bibinfo {author} {\bibfnamefont {E.~G.}\ \bibnamefont {Sorte}}, \
  and\ \bibinfo {author} {\bibfnamefont {B.}~\bibnamefont {Saam}},\ }\href
  {\doibase 10.1103/PhysRevLett.106.193005} {\bibfield  {journal} {\bibinfo
  {journal} {Phys. Rev. Lett.}\ }\textbf {\bibinfo {volume} {106}},\ \bibinfo
  {pages} {193005} (\bibinfo {year} {2011})}\BibitemShut {NoStop}%
\bibitem [{\citenamefont {Korver}\ \emph {et~al.}(2013)\citenamefont {Korver},
  \citenamefont {Wyllie}, \citenamefont {Lancor},\ and\ \citenamefont
  {Walker}}]{Korver2013}%
  \BibitemOpen
  \bibfield  {author} {\bibinfo {author} {\bibfnamefont {A.}~\bibnamefont
  {Korver}}, \bibinfo {author} {\bibfnamefont {R.}~\bibnamefont {Wyllie}},
  \bibinfo {author} {\bibfnamefont {B.}~\bibnamefont {Lancor}}, \ and\ \bibinfo
  {author} {\bibfnamefont {T.~G.}\ \bibnamefont {Walker}},\ }\href {\doibase
  10.1103/PhysRevLett.111.043002} {\bibfield  {journal} {\bibinfo  {journal}
  {Phys. Rev. Lett.}\ }\textbf {\bibinfo {volume} {111}},\ \bibinfo {pages}
  {043002} (\bibinfo {year} {2013})}\BibitemShut {NoStop}%
\bibitem [{\citenamefont {Bell}\ and\ \citenamefont {Bloom}(1961)}]{Bell1961}%
  \BibitemOpen
  \bibfield  {author} {\bibinfo {author} {\bibfnamefont {W.~E.}\ \bibnamefont
  {Bell}}\ and\ \bibinfo {author} {\bibfnamefont {A.~L.}\ \bibnamefont
  {Bloom}},\ }\href {\doibase 10.1103/PhysRevLett.6.280} {\bibfield  {journal}
  {\bibinfo  {journal} {Phys. Rev. Lett.}\ }\textbf {\bibinfo {volume} {6}},\
  \bibinfo {pages} {280} (\bibinfo {year} {1961})}\BibitemShut {NoStop}%
\bibitem [{\citenamefont {Kornack}\ and\ \citenamefont
  {Romalis}(2002)}]{Kornack2002}%
  \BibitemOpen
  \bibfield  {author} {\bibinfo {author} {\bibfnamefont {T.~W.}\ \bibnamefont
  {Kornack}}\ and\ \bibinfo {author} {\bibfnamefont {M.~V.}\ \bibnamefont
  {Romalis}},\ }\href {\doibase 10.1103/PhysRevLett.89.253002} {\bibfield
  {journal} {\bibinfo  {journal} {Phys. Rev. Lett.}\ }\textbf {\bibinfo
  {volume} {89}},\ \bibinfo {pages} {253002} (\bibinfo {year}
  {2002})}\BibitemShut {NoStop}%
\end{thebibliography}

%

\end{document}